# Enhanced valley splitting in monolayer WSe$_2$ due to magnetic exchange field

Chuan Zhao[1], Tenzin Norden[1], Puqin Zhao[2,1], Yingchun Cheng[2], Peiyao Zhang[1], Fan Sun[1],

Payam Taheri[1], Jieqiong Wang[3], Yihang Yang[4], Thomas Scrace[1], Kaifei Kang[3,1], Sen Yang[3],

Guo-xing Miao[4], Renat Sabirianov[5], George Kioseoglou[6], Athos Petrou[1], and Hao Zeng[1]

[1]Department of Physics, University at Buffalo, the State University of New York, Buffalo, NY 14260, USA;

[2]Key Laboratory of Flexible Electronics (KLOFE) & Institute of Advanced Materials (IAM), Jiangsu National Synergetic Innovation Center for Advanced Materials (SICAM), Nanjing Tech University, Nanjing, China;

[3]College of Science, Xi'an Jiaotong University, Xi'an, China;

[4]Institute of Quantum Computing, University of Waterloo, Waterloo, ON, Canada;

[5]Department of Physics, University of Nebraska-Omaha, Omaha, NE 68182, USA;

[6]Department of Material Science and Technology, University of Crete, Heraklion, GR 71003, Greece

**Exploiting the valley degree of freedom to store and manipulate information provides a novel paradigm for future electronics. A monolayer transition metal dichalcogenide (TMDC) with broken inversion symmetry possesses two degenerate yet inequivalent valleys[1,2], offering unique opportunities for valley control through helicity of light[3-5]. Lifting the valley degeneracy by Zeeman splitting has been demonstrated recently, which may enable valley control by a magnetic field [6-9]. However, the realized valley splitting is modest, (~ 0.2 meV/T). Here we show greatly enhanced valley spitting in monolayer $WSe_2$, utilizing the interfacial magnetic exchange field (MEF) from a ferromagnetic EuS substrate. A valley splitting of 2.5 meV is demonstrated at 1 T by magneto-reflectance measurements. Moreover, the splitting follows the magnetization of EuS, a hallmark of the MEF. Utilizing MEF of a magnetic insulator can induce magnetic order, and valley and spin polarization in TMDCs, which may enable valleytronic and quantum computing applications[10-12].**

TMDCs such as $MoS(Se)_2$ and $WS(Se)_2$ are semiconducting graphite analogues composed of a layer of atoms covalently bonded, and stacks of these layers held together by van der Waals interactions [13,14]. Monolayer TMDCs with broken inversion symmetry possesses two degenerate yet inequivalent valleys, related by time-reversal symmetry [2]. This property and strong spin-orbit coupling is responsible for the unique physics of TMDCs such as coupled spin and valley degrees of freedom [1]. Together with a direct band gap, TMDCs offer the opportunity to selectively excite carriers within a particular valley with specific valley pseudospin using circularly polarized light [3-5,15,16]. Furthermore, in electron or hole doped samples, valley Hall and spin Hall effects can be observed [1,17,18]. Lifting the valley degeneracy in these materials is of great interest because it would allow for control of valley polarization for memory and logic applications [17,19,20]. This has been achieved recently by applying an external magnetic field to Zeeman split the band edge states in different valleys [7-9]. The valley splitting reported, however, is small (~ 0.2 meV/T), making magnetic control only feasible at high fields and practical applications difficult [21]. It has also been reported that valley degeneracy can be broken by intense circularly polarized light through the optical stark effect [22,23].

An alternative approach to overcome the small valley splitting issue is to utilize an interfacial MEF [10,11], which breaks the time reversal symmetry. A giant and tunable valley splitting has been theoretically predicted in monolayer $MoTe_2$ on a EuO substrate [10,11]. The exchange coupling between the ferromagnetic EuO and $MoTe_2$ can result in a valley splitting of 44 meV [11]. Experimentally, EuS is among the few known magnetic insulators chosen to provide the MEF due to the large magnetic moment of $Eu^{2+}$ ($\langle S_z \rangle$ ~ 7 $\mu_B$) and large exchange coupling (J ~ 10 meV), leading to a large MEF $\propto J\langle S_z \rangle$ [24]. Recently, a substantial MEF of 14 T has been measured by Zeeman spin Hall effect in Graphene/EuS heterostructures [25]. Proximity-induced ferromagnetism in graphene and topological insulators have also been reported [26,27].

In this work, we demonstrate experimentally greatly enhanced valley splitting in monolayer $WSe_2$, using the MEF induced by an EuS substrate. Its valley specific optical inter-band transitions were measured by magneto-reflectance to probe the exciton valley splitting. The valley splitting of $WSe_2$/EuS samples has been enhanced by an order of magnitude to 2.5 meV at 1 T. More importantly, the field-dependence of Zeeman splitting follows the magnetic hysteresis of EuS, a hallmark of exchange-field induced Zeeman splitting. Our work shows that harnessing the MEF from a ferromagnetic material is an effective approach for valley control and inducing valley/spin polarization in monolayer TMDCs, which can be superior to spin injection with potential difficulties such as pinholes and barrier breakdown.

Magneto-reflectance measurements were carried out in the Faraday geometry (see schematics S4 in Supplemental Information, SI) on monolayer WSe$_2$ in order to determine their excitonic transitions and from these the valley splitting. In our study, we focused only on the lowest energy "A" exciton transitions. In Fig. 1 we compare the reflectance spectra of monolayer WSe$_2$ on both SiO$_2$ and EuS substrates, measured at 0 and ±7 T, at 7 K. The vertical axis is the ratio $R/R_0$, where $R$ denotes the reflectance from WSe$_2$ and $R_0$ denotes the reflectance from the SiO$_2$ or EuS substrate adjacent to WSe$_2$. The middle plot in Fig. 1(a) shows the zero field reflectance signal from the "A" exciton of monolayer WSe$_2$ on SiO$_2$; left-circularly polarized light ($\sigma_+$) corresponds to the K and right-circularly polarized light ($\sigma_-$) corresponds to the K' valleys, respectively. The exciton features were fitted using complex (absorptive + dispersive) Fano line-shape to extract the transition energies. A clear transition at 1.76 eV is observed at $B = 0$, which is consistent with earlier reported value for "A" exciton of monolayer WSe$_2$ [28]. As expected, the $\sigma_+$ and $\sigma_-$ spectra match perfectly with each other, indicating no energy splitting of the two valleys as required by the time-reversal symmetry. Similarly, there is no measurable excitonic splitting at $B = 0$ for WSe$_2$ on EuS, as shown in the middle plot of Fig. 1(b). Upon application of a +7 T magnetic field, however, a clear valley splitting is observed for WSe$_2$ on SiO$_2$ (Top plot of Fig. 1(a)). The $\sigma_+$ spectrum ("A" exciton transition at the K valley) shifts to lower energy, while the $\sigma_-$ spectrum shifts to higher energy. The valley splitting is defined as $\Delta E \equiv E(\sigma_-) - E(\sigma_+)$, where $E(\sigma_+)$ and $E(\sigma_-)$ refer to the fitted peak energy of $\sigma_+$ and $\sigma_-$, respectively. When a -7 T field is applied, the sign of the valley splitting is reversed, as shown in the lower plot of Fig. 1(a); $\Delta E$ is 1.5 meV at 7 T, which is consistent with earlier reported values[7-9]. In WSe$_2$ on EuS the valley splitting is equal to 3.9 meV at 7 T (see Fig. 1(b)). This enhanced value implies that the valley splitting does not just come from the Zeeman effect due to the external field, but there is an important contribution from the ferromagnetic EuS substrate.

To elucidate the role of EuS, the field-dependence of valley Zeeman splitting $\Delta E$ measured at 7 K is shown in Fig. 2. For $WSe_2$ on a $SiO_2$ substrate, the field dependence is linear, with a slope of 0.20 meV/T, consistent with earlier reported values [9,21]. In contrast, $\Delta E$ for $WSe_2$ on an EuS substrate shows pronounced nonlinear behavior: it first increases rapidly with increasing field for -1 T< $B$ < +1 T. For $|B| > 1$ T, the slope decreases with increasing field and eventually reaches a constant value of 0.20 meV/T, very close to the slope for $WSe_2$ on a $SiO_2$ substrate. The slope for -1 T< $B$ < +1 T, however, is 2.5 meV/T, which is an order of magnitude higher than the slope for $WSe_2$ on a $SiO_2$ substrate. The enhanced valley splitting value in $WSe_2$/EuS suggests an effective exchange field of about 12 T at $B = 1$T possibly resulting from the interfacial MEF.

Below we discuss how spurious effects, in particular the magnetic dipole field contribution from EuS, are ruled out. We estimate the maximum dipole field contribution from EuS to be 1.2 T (see discussion below), which is an order of magnitude smaller than the exchange field. Thus the dipole field of EuS cannot be a significant contributor to the enhanced valley splitting. Furthermore, if the enhanced valley splitting is indeed due to the interfacial exchange field, inserting a non-magnetic spacer layer between $WSe_2$ and EuS should interrupt the exchange coupling due to its short-range nature. To check this, we measured the valley splitting of $WSe_2$ on EuS with a 10 nm $SiO_2$ spacer inserted in between them. In Fig. 2(b) we plot the field dependent valley splitting for the $WSe_2$/EuS and $WSe_2$/$SiO_2$/EuS samples. As can be seen from Fig. 2(b), there is no measureable difference between the behavior of the exciton valley splittings of $WSe_2$ on EuS with the $SiO_2$ spacer and that of $WSe_2$ on the $SiO_2$ substrate. This clearly rules out the dipole field as the origin of the enhanced valley splitting, since the stray field is not significantly reduced by the 10 nm spacer.

The interfacial MEF should be proportional to the magnetization of the ferromagnetic substrate. To unambiguously confirm the origin of the enhanced valley splitting, systematic field and temperature dependence of the splitting were measured (Fig. 3(a)), and compared to the magnetic hysteresis loops of EuS measured at different temperatures (Fig. 3(b)). A linear background of 0.20 meV/T is subtracted from all curves in Fig. 3(a), and the net value is attributed to MEF and denoted as $\Delta E_{ex}$. It can be seen that at 7, 12 and 20 K, $\Delta E_{ex}$ shows non-linear field dependence: at low fields, $\Delta E_{ex}$ increases rapidly with increasing field, starting from an initial value close to zero at $B = 0$. With further increasing field, the rate of increase of $\Delta E_{ex}$ slows down and then tends toward saturation at high fields. With increasing temperature, $\Delta E_{ex}$ decreases accordingly. Such behaviors are very similar to the field and temperature dependent magnetization of EuS, as shown in Fig. 3(b). The measured saturation magnetization $M_S$ of EuS at 7 K is 1,000 emu/cm$^3$, slightly lower than bulk $M_S$ value of 1,200 emu/cm$^3$ (~ 7 $\mu_B$/Eu$^{2+}$) at 0 K. EuS has a cubic structure and is magnetically soft. The strong shape anisotropy of the 10 nm thin film forces the magnetic easy axis to lie in the plane. This explains the shape of the magnetic hysteresis loops with very low remanent magnetization, and a relatively high saturation field (~ 1.2 T), corresponding to the demagnetization field of $4\pi M_S$. Therefore, WSe$_2$ does not show measurable spontaneous valley splitting even below the EuS Curie temperature ($T_C$) of 16.6 K. Similar behavior is also observed in the PL polarization in spin-LEDs with an Fe spin injector[29]. It should be noted that $\Delta E_{ex}$ vs $B$ shows non-linear behavior even at 20 K, above expected $T_C$ of EuS. This is also consistent with temperature dependent magnetization showing a tail well above 20 K (Fig. S5, SI). The magnitude of $\Delta E_{ex}$ at 7 T is comparable to the Zeeman splitting due to the external field alone. At 50 $K$, $\Delta E_{ex}$ value is greatly reduced and it increases with $B$ essentially linearly with a slope of just 0.04 meV/T as shown in Fig. 3(a), which is expected from the paramagnetic behavior of EuS. The very similar field and temperature dependence shown in Figs. 3(a) and 3(b) suggests that $\Delta E_{ex}$ is directly correlated to the out-of-plane magnetization of EuS. To demonstrate further the correlation, Fig. 3(c) shows $\Delta E_{ex}$ vs $B$ measured at different temperatures superimposed on magnetization $M$ vs $B$. Both $\Delta E_{ex}$ and $M$ values are normalized by their respective saturated values at 7 K. Remarkably, the valley exchange splitting $\Delta E_{ex}$ of WSe$_2$ measured by magneto-reflectance and the out-of-plane magnetization $M$ of EuS measured by a vibrating sample magnetometer (VSM) match with each other well, for the measurement temperatures of 7 and 12 K. This shows unambiguously that the valley exciton splitting in monolayer WSe$_2$ originates from the interfacial MEF, and thus scales with the magnetization of EuS. For higher temperatures, there is some noticeable deviation. While the exact reason is not clear, we cannot rule out the possibility of a small shift of the sample spot while the temperature was raised.

We further investigate the valley exchange splitting and magnitude of MEF using Density Functional calculations of band structure of a monolayer of WSe$_2$ deposited on EuS(111) slab substrate. The system is shown schematically in Fig. 4(a). There is a slight mismatch in lattice parameters of EuS and WSe$_2$ of about ~3% with the unit cell of $\vec{a'} = 3\vec{a} - \vec{b}$ ($\vec{a}, \vec{b}$ are primitive lattice vectors of WSe$_2$) as shown in Fig. 4(b). Figure 4(c) shows the band structure of WSe$_2$/EuS bi-layer in the vicinity of the band gap along Γ-K and Γ-K' directions calculated by DFT. The direct optical band gap occurs at the K (K') point. Focusing now on the top valence and bottom conduction bands, as shown schematically in Fig. 4(d). The top two valence bands in WSe$_2$ are spin split at the K point by ~0.46 eV, due to the strong spin-orbit coupling on W site. The highest valence band at K point has spin up character (red), while at K' it is spin down (blue). Opposite is true for the second highest valence band because of opposite spin of the electron in this band with respect to the top valence band. The conduction bands show a much smaller spin splitting (~0.041 eV) due to the spin orbit coupling. In the absence of a magnetic field, the energy of the top valence (bottom conduction) bands (shown by the dashed lines) at K and K' are degenerate due to the time-reversal symmetry, despite different spin characters.

The degeneracy is lifted, however, when $WSe_2$ is subjected to an external magnetic field or exchange field due to the EuS substrate. Bands of opposite spin characters are shifted in opposite directions, *i.e.* spin-up bands (red) are shifted downward and spin-down bands (blue) are shifted upward. The shifted bands are shown by the solid lines in Fig. 4(d). With an external magnetic field, the energy level shifts can be attributed to contributions from the spin, atomic orbital and valley orbital magnetic moments.[6-9] The energy shift due to these different contributions are indicated by black and green arrows in Fig. 4(d). The spin magnetic moment should not affect the optical transition as it shifts both conduction and valence bands by the same amount. The valley orbital moment contribution is also negligible due to small differences in effective mass [7,21]. The atomic orbital moment, on the other hand, differs for valence and conduction bands because conduction band is mainly composed of *d*-orbital with magnetic quantum number $m = 0$, while the valence band corresponds to *d*-orbitals with $m = 2$ in K and $m = -2$ in K' valleys, respectfully [7]. It does clearly contribute to the energy splitting of valley exciton transitions. In the presence of exchange fields, we expect dominating contributions to the valley exciton splitting from atomic orbital moment. However, the spin moment contribution may not be ruled out. Due to the difference in the symmetry of *d*-orbitals, the Zeeman-like exchange contribution from the spin magnetic moment to the conduction and valence bands can be different. They are opposite in K and K' valleys, which can then contribute to the valley exciton splitting. As a result, the inter-band transition ("A" exciton) at the K valley for spin up bands (σ+) is $\Delta_{opt}^{\uparrow}(K) = E(c, K) - E(v, K) = 1.106$ eV, *vs* that at K' valley for spin down bands (σ-) $\Delta_{opt}^{\downarrow}(K') = E(c, K') - E(v, K') = 1.116$ eV (Table I in SI). The energy splitting for the "A" exciton at K and K' valleys is thus about 10 meV, equivalent to an external magnetic field of about 50 T. Due to the large magnitude of the MEF, EuS serves as a "magnetic field amplifier" to enhance the valley exciton splitting.

The experimental splitting is a few times smaller than theoretically predicted value. This is not surprising considering that the EuS surface is modeled as an ideal Eu-terminated surface, while experimentally prepared interface between WSe$_2$ and EuS is more complex. EuS grown by e-beam evaporation is polycrystalline with only a fraction of surface-reconstructed (111) facets exposed. The other high symmetry surfaces such as (100) and (110) have lattice mismatch and considerably less Eu sites at the surface. However, it is clear that by optimizing the interface, there is great room to enhance the experimental valley splitting. By using a magnetic insulator with $T_C$ above room temperature such as Yttrium Iron Garnet, it is also possible to realize enhanced valley splitting at room temperature, which is critical for device applications.

In conclusion, we demonstrated greatly enhanced valley exciton splitting in monolayer WSe$_2$ utilizing the interfacial MEF from the ferromagnetic EuS substrate. The valley splitting is enhanced by more than an order of magnitude, equivalent to an effective magnetic field of 12 T. The field and temperature dependence of splitting scales with the magnetization of EuS, confirming the exchange field origin. Our work offers enhanced capability to control valley and spin polarization. For example, by electric gating, it is possible to tune the chemical potential to polarize selected valleys. Since the charge carriers are also carriers of spin and valley-dependent orbital angular momentum, anomalous charge, spin, and valley Hall effects are expected. The convenient manipulation of such degrees of freedom offers new paradigm for classical and quantum information processing applications.

**Acknowledgement:**

Work supported by US National Science Foundation MRI-1229208, DMR-1104994，DMR-1305770, CBET-1510121, the Natural Sciences and Engineering Research Council of Canada (NSERC) Discovery grant RGPIN 418415-2012 and the National Natural Science Foundation of China (Nos. 11504169, and 61575094).

**Author contributions**

C.Z., P.-Q.Z., P.T., K.K. and J.W. prepared and characterized the monolayer TMDCs including WSe$_2$ and transferred them onto EuS substrates. T.S., P.Z., T.N., C.Z. and A.P. performed magneto-optical measurements and data analysis. Y.Y. and G.M. provided the EuS thin films. F.S. performed magnetic measurements of EuS. R.S. and Y.C. performed the first-principle calculations. H.Z., C.Z., Y.C., R.S. G.K. and A.P. wrote the manuscript. H.Z. and A.P. guided the project. All authors commented on the manuscript.

**Figure captions:**

**Figure 1** Reflectance spectra from "A" exciton of monolayer WSe$_2$ recorded at $T$ =7 K. (a) WSe$_2$ on Si/SiO$_2$ substrate and (b) WSe$_2$ on EuS substrate. Top: $B$ =+7 T, Middle: $B$ =0 and Bottom: $B$ =-7 T. $\sigma_+$ ($\sigma_-$) corresponds to the transition at K (K') valley. There is no splitting for spectra at $B$ =0, for either the Si/SiO$_2$ or the EuS substrate. At +7 T, the $\sigma_+$ ($\sigma_-$) component shifts to lower (higher) energy. At -7 T, the energy shift is in the opposite direction. By comparing Fig. 1(a) and 1(b), it is clear that the energy splitting for WSe$_2$ on the EuS substrate is noticeably higher than that on Si/SiO$_2$ substrate. The red circles (blue squares) indicate $\sigma_+$ ($\sigma_-$) incident polarization. The dots are experimental data and solid lines are the fitting results using "absorptive and dispersive" line-shape.

**Figure 2** Measured valley splitting $\Delta E$ as a function of magnetic Field. **(a)** WSe$_2$ on EuS vs SiO$_2$ substrates. Purple circles represent the data on the Si/SiO$_2$ substrate and green squares are data on the EuS substrate. On Si/SiO$_2$, with increasing field, $\Delta E$ increases linearly with a slope of 0.20 meV/T. On the EuS substrate, the splitting is greatly enhanced, with an initial slope of 2.5 meV/T at -1 T< $B$ < +1 T. At |$B$| larger than 1 T, the slope gradually decreases to the value of WSe$_2$ on the Si/SiO$_2$ substrate. **(b)** WSe$_2$ on SiO$_2$ (10 nm)/EuS vs SiO$_2$ substrates. Purple circles are the data on the Si/SiO$_2$ while yellow squares represent the data on 10nm SiO$_2$ spacer. Both show linear field dependence with identical slope of 0.20 meV/T.

**Figure 3** Comparing magnetic field-dependent valley exchange splitting of WSe$_2$/EuS and magnetic hysteresis loops of EuS measured at different temperatures. **(a)** Field-dependent valley exchange splitting $\Delta E_{ex}$ due exclusively to MEF for WSe$_2$ measured at 7, 12, 20 and 50 K, respectively. A linear background of 0.20 meV/T, attributed to the Zeeman splitting due to external $B$ field is subtracted. $\Delta E_{ex}$ shows non-linear field-dependence at 7 K (black squares), 12 K (red circles) and 20 K (blue up-triangles), but a linear field-dependence at 50 K (purple down-triangles); **(b)** Magnetization $M$ of EuS as a function of field measured at 7, 12, 20 and 50 K, respectively. The saturation magnetization at 7 K is 1,000

emu/cm$^3$, slightly lower than 1,200 emu/cm$^3$ expected at 0 K. The similarity between $M$ and $\Delta E_{ex}$ is apparent; (**c**) Field-dependent $\Delta E_{ex}$ of WSe$_2$ and $M$ of EuS superimposed on top of each other. Both $\Delta E_{ex}$ and $M$ are normalized by their saturated values at 7 K. Dots represent normalized $\Delta E_{ex}$ and lines represent normalized $M$. It can be seen that they match with each other well at 7 and 12 K, confirming unambiguously that the origin of the enhanced valley splitting is due to the magnetic exchange field. They show deviations from each other at higher temperatures possibly due to a slight shift in sample position in magneto-reflectance measurements.

**Figure 4** Calculated band structure and valley exchange splitting of WSe$_2$/EuS. (a) Schematic diagram of WSe$_2$ monolayer deposited on the ferromagnetic EuS substrate with magnetization perpendicular to the plane. (b) Top view of WSe$_2$/EuS interface used in calculations. The lattice parameters of the selected unit cell of WSe$_2$ matches closely to that of EuS. (c) The band structure of WSe$_2$/EuS bi-layer in the vicinity of the band gap along Γ-K and Γ-K' directions calculated by DFT. (d) A schematic energy diagram at the K and K' valleys showing energy splitting of the bands in exchange coupled WSe$_2$/EuS, and the corresponding "A" exciton transitions at the K and K' valleys. Spin-up bands are represented by red and spin-down bands are represented by blue colors. Black arrows represent the net energy shift due to spin and valley orbital moment contributions. Green arrows represent the energy shift due to atomic orbital moment contributions. The valley exciton splitting originates mainly from the atomic orbital moment contributions. Spin moment contribution cannot be ruled out due to different exchange coupling to the conduction and valence bands.

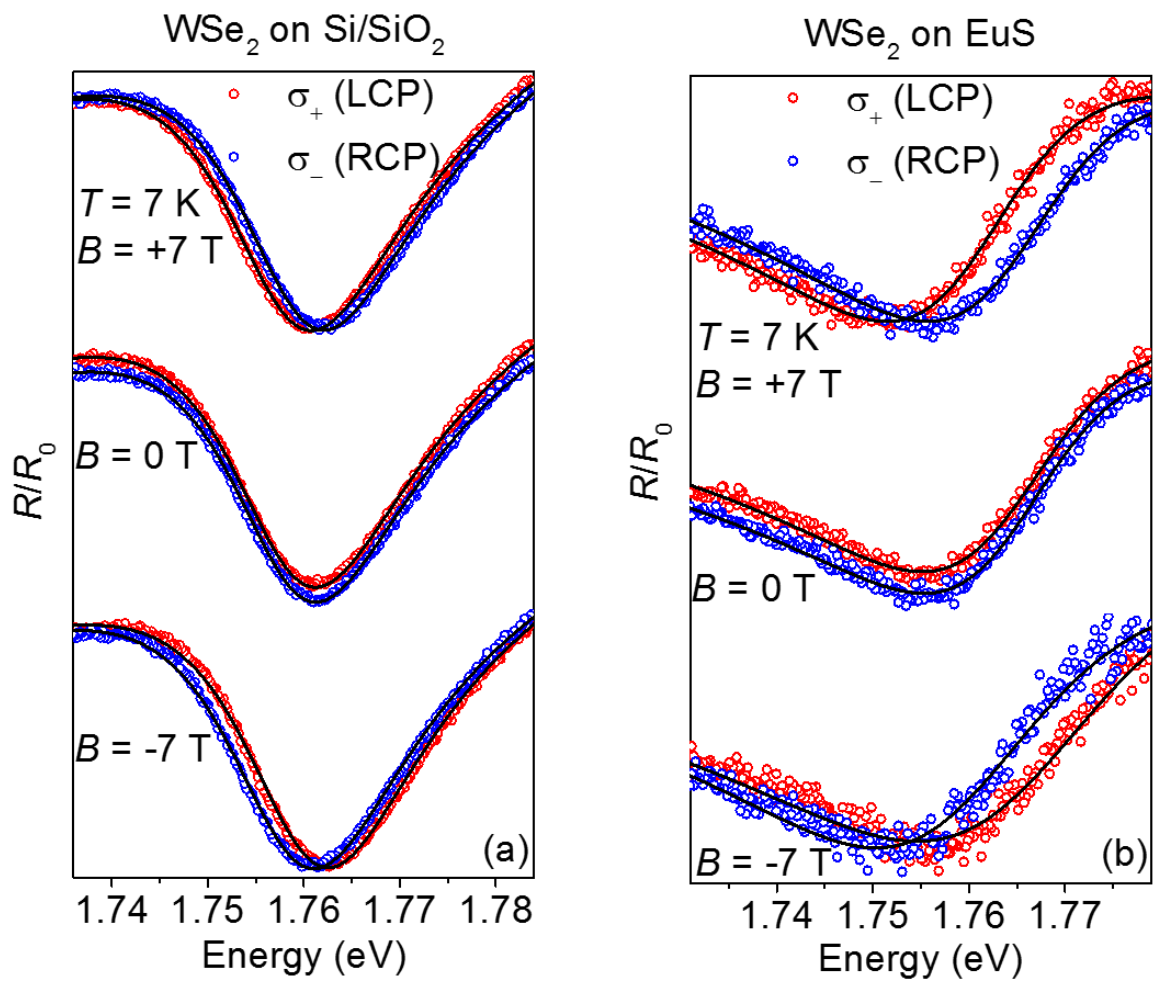

**Figure 1**

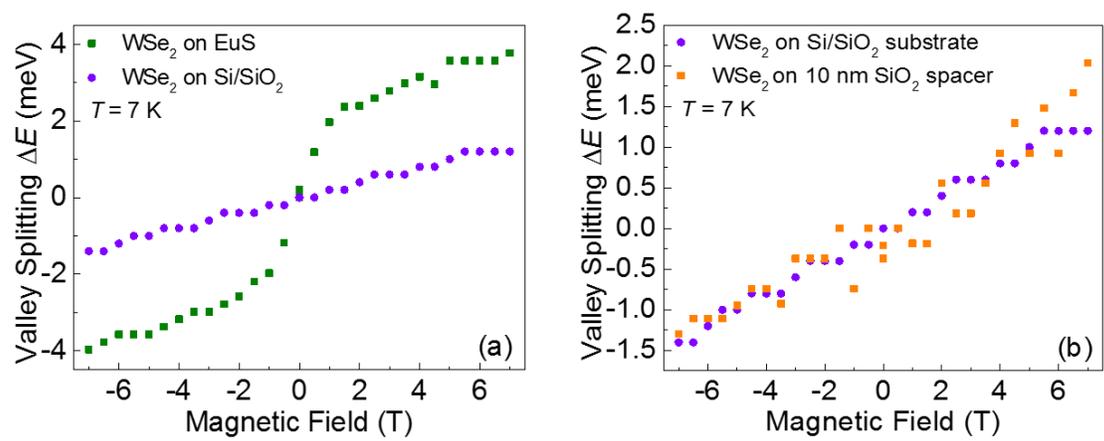

**Figure 2**

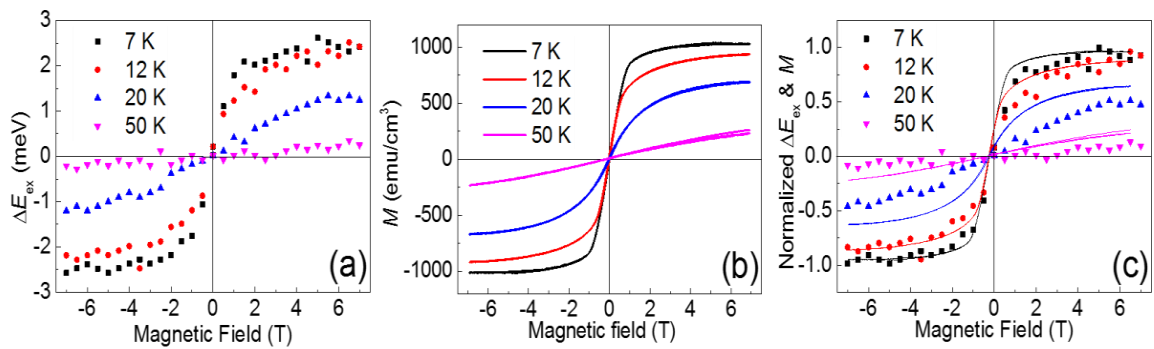

**Figure 3**

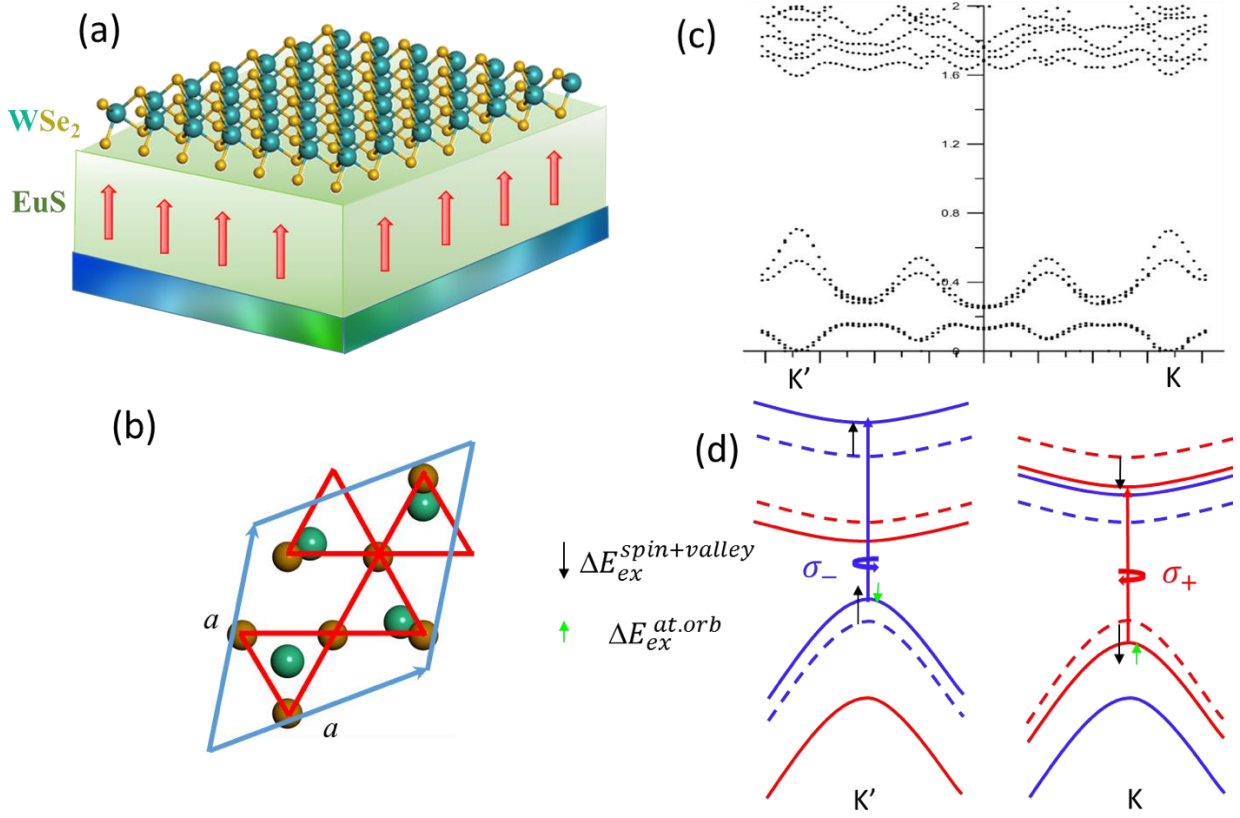

**Figure 4**

**Methods**

Monolayer TMDCs including WSe$_2$ were prepared by selenization of electron-beam evaporated ultrathin transition metal oxide films on sapphire substrates, similar to our previous work on MoS$_2$ as described in reference 31 and SI. [31]

For reflectivity measurements, the as-grown monolayer TMDC films were transferred onto Si/SiO$_2$ and EuS substrates, modified from published procedures[32]. Briefly, monolayer WSe$_2$ on sapphire substrate was covered by PMMA by spin coating. After 5 min baking at 50 °C, a water droplet was placed on PMMA surface. The sample's edge was then poked by tweezers, and the water penetrated between the film and substrate. After 10 min, the film completely separated from the substrate and floated on the water surface. The film was then transferred onto Si/SiO$_2$ or EuS substrate, followed by 5 min baking at 80 °C. Baking of PMMA at sufficiently high temperatures stretches the PMMA, and helps to eliminate wrinkles in TMDCs. The PMMA was removed by immersing the sample in acetone for 5 min. After repeated cleaning in acetone, the sample was then annealed in an ultrahigh vacuum chamber at 350 °C for 30 min to remove any potential adsorbates and improve the interface quality.

For magneto-reflectance measurements, the samples were placed on the cold finger of a continuous flow optical cryostat operated in the 5-300 K temperature range. The cryostat was mounted on a three-axis translator with a spatial resolution of 10 μm in each direction. The *x*- and *y*-translation stages allow us to access a single TMDC crystal. The cryostat tail was positioned inside the room temperature bore of a 7 tesla superconducting magnet. A collimated white light beam was used for the reflectivity work. The incident light was focused on the sample using a microscope objective with a working distance of 34 mm. The incident beam was polarized either as left-circularly polarized (LCP, $\sigma_+$) or right circularly polarized (RCP, $\sigma_-$) using a Babinet-Soleil compensator. The objective collected the reflected beam from the sample in the Faraday geometry and the light was focused onto the entrance slit of a single monochromator that uses a cooled charge coupled device (CCD) detector array.

The temperature dependent magnetic hysteresis loops of EuS were measured by the vibrating sample magnetometer (VSM) option of a Quantum Design Physical Property Measurement System. The magnetic field was applied in the direction perpendicular to the film plane, and thus only the out-of-plane component of the magnetic moment was measured.

**Supplemental Information**

**Enhanced valley splitting in monolayer WSe$_2$ due to magnetic exchange field**

Chuan Zhao[1], Tenzin Norden[1], Puqin Zhao[2], Yingchun Cheng[2], Peiyao Zhang[1], Fan Sun[1], Payam Taheri[1], Jieqiong Wang[3], Yihang Yang[4], Thomas Scrace[1], Kaifei Kang[3,1], Sen Yang[3], Guoxing Miao[4], Renat Sabirianov[5], George Kioseoglou[6], Athos Petrou[1], and Hao Zeng[1]

[1]Department of Physics, University at Buffalo, the State University of New York, Buffalo, NY 14260, USA;

[2]Key Laboratory of Flexible Electronics (KLOFE) & Institute of Advanced Materials (IAM), Jiangsu National Synergetic Innovation Center for Advanced Materials (SICAM), Nanjing Tech University, Nanjing, China;

[3]College of Science, Xi'an Jiaotong University, Xi'an, China;

[4]Institute of Quantum Computing, University of Waterloo, Waterloo, ON, Canada;

[5]Department of Physics, University of Nebraska-Omaha, Omaha, NE 68182, USA;

[6]Department of Material Science and Technology, University of Crete, Heraklion, GR 71003, Greece

**Material synthesis**

In a typical synthesis of monolayer WSe$_2$, a WO$_3$ film with a thickness of 6-10 Å was deposited on a sapphire (0001) substrate. The oxide film was then placed in a two-zone tube furnace under N$_2$ flow, with selenium powder located on the upstream, as shown in Fig. S1(a). The WO$_3$ film reacts with selenium at high temperatures to form WSe$_2$. By controlling the synthetic conditions including the WO$_3$ film thickness, the amount of selenium source, flow rate and heating profile, monolayer WSe$_2$ nearly free of overgrowth can be achieved. A typical heating profile of the surface for monolayer WSe$_2$ growth is shown in Fig. S1(b).

10 nm polycrystalline EuS thin films were grown by electron-beam evaporation on Si/SiO$_2$ substrates at room temperature.

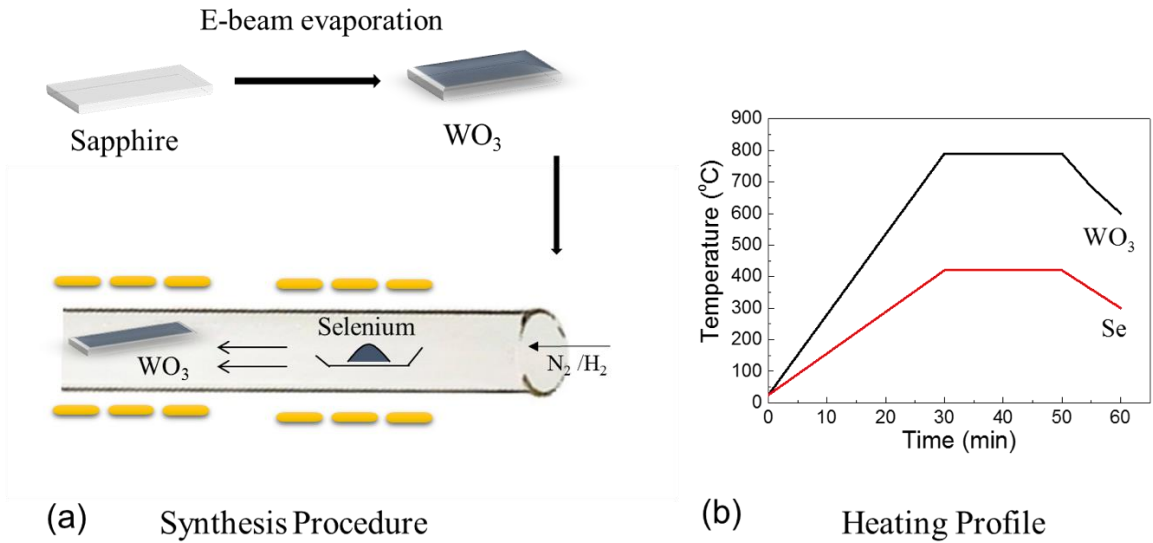

**Figure S1** (a) The experimental setup for the synthesis of WSe$_2$ monolayer and (b) the heating profiles for the two-zone furnace in WSe$_2$ synthesis.

**Structural characterization of monolayer WSe$_2$**

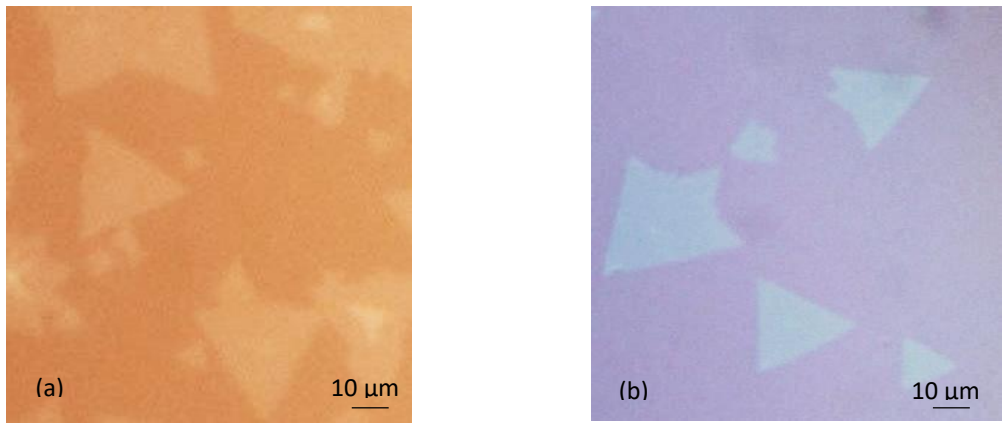

**Figure S2** Optical microscope images of (a) monolayer WSe$_2$ on sapphire substrate and (b) monolayer WSe$_2$ on EuS. In Fig. S2(a), the as-grown WSe$_2$ films consist of discrete, triangular shaped single crystal monolayers on sapphire substrate, with sizes of 10-20 micrometers. Figure S2(b) is the corresponding image of WSe$_2$ films after being transferred onto EuS substrate. It can be seen that the transfer does not result in change in morphology. The samples are nearly free of residues or wrinkles.

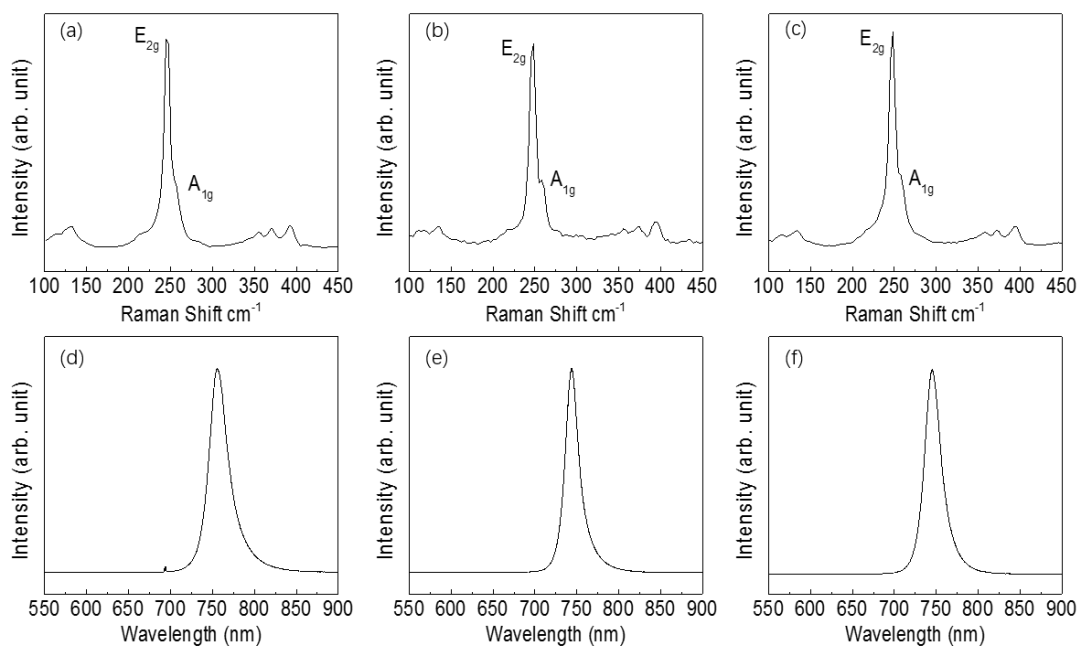

**Figure S3** Room temperature Raman and PL spectra of WSe$_2$ on different substrates. Figure S3(a) shows the Raman spectrum of monolayer WSe$_2$ on sapphire substrate. The E$_{2g}$ and A$^1_g$ peaks are located at 247.3 and 257.6 cm$^{-1}$, respectively, with a peak separation of 10.3 cm$^{-1}$, consistent with reported value for monolayer WSe$_2$ [30]. After being transferred onto SiO$_2$ (S3(b)) and EuS (S3(c)) substrates, the peak positions remain nearly unchanged. The PL peak energy for sample on sapphire is at 1.64 eV (756 nm), as shown in Fig. S3(d), which is attributed to the "A" excition transition. The PL peak energy values change slightly to 1.66 eV (745 nm) and 1.67 eV (744 nm) for samples on SiO$_2$ and EuS substrates, as shown in Fig. S3(e) and S3(f), respectively. These values are consistent with values reported previously for monolayer WSe$_2$ [30].

**Magneto-reflectance measurements**

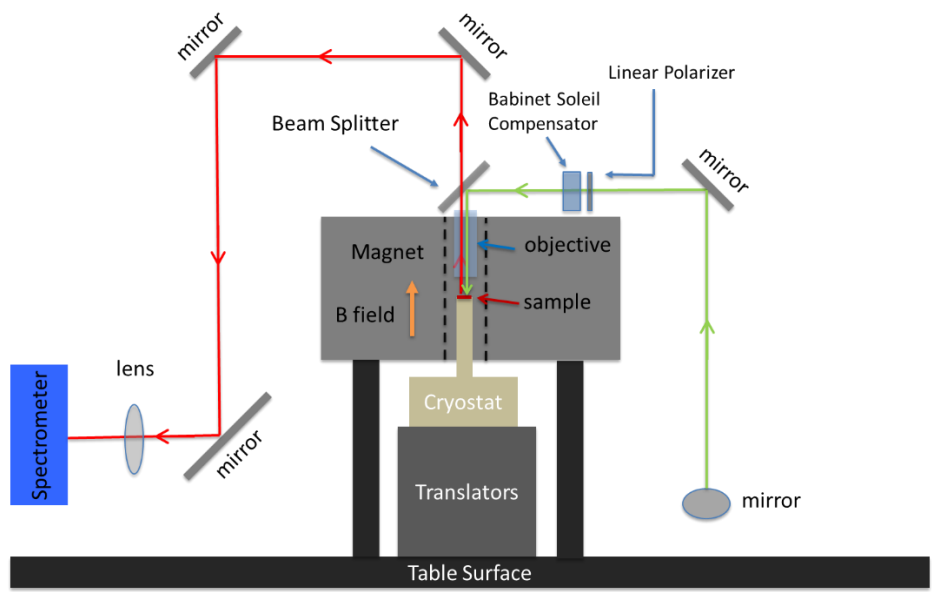

**Figure S4** A schematic of the setup for magneto-reflectance measurements. Positive magnetic field is defined as the upward direction. The light from the sample is analyzed in its right and left circularly polarized components by a combination of a quarter wave plate and a linear polarizer.

**Magnetization as a function of temperature for EuS**

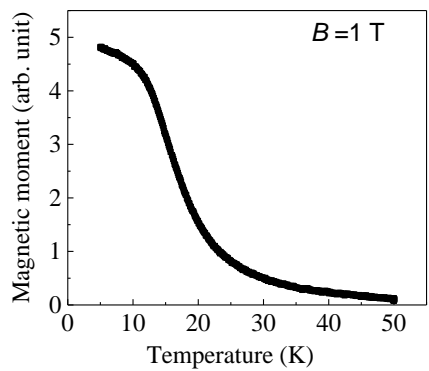

**Figure S5** The magnetic moment of 10 nm EuS as a function of temperature measured at 1 T.

**Band structure calculations of WSe₂/EuS**

Our spin-polarized *ab initio* calculations for WSe$_2$/EuS are based on density functional theory (DFT) [1,2] and within the generalized gradient approximation (GGA) formulated by Perdew, Burke, and Ernzerhof (PBE) to the exchange-correlation energy functional [3]. The Kohn-Sham equations are solved employing the all-electron projected augmented wave PAW method, as implemented in the Vienna *ab initio* simulation package VASP [4]. We considered all PAW projectors for which the semicore electrons are taken into account as valence electrons [5]. The cutoff energy value of 400 eV is used. For all calculations, the equilibrium geometries are obtained when the atomic forces are smaller than 0.01 eV/Å and the total energy converges within $10^{-6}$ eV. EuS magnetized out of plane was considered.

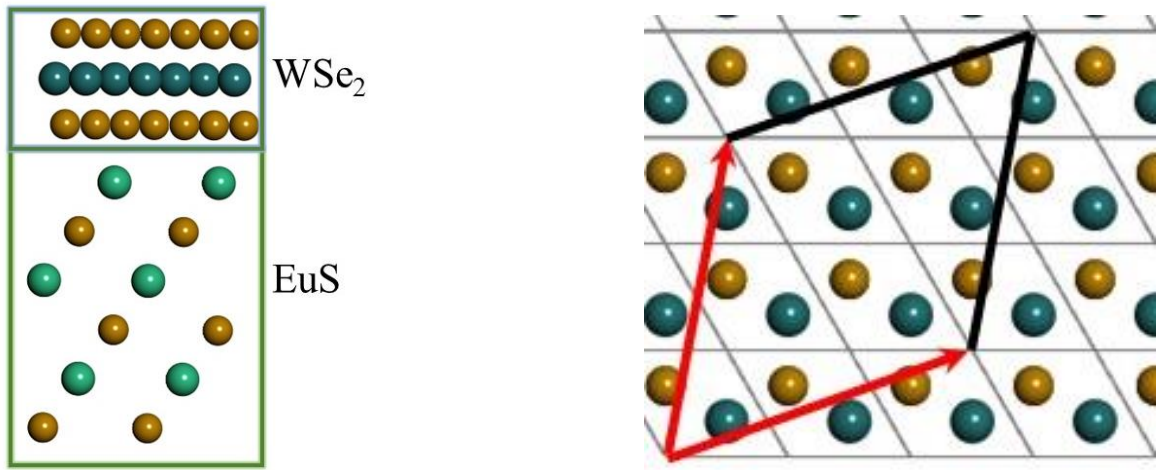

**Figure S6.** a) Model WSe$_2$ monolayer placed on the (111) surface of EuS (terminated by Eu). b) The lattice parameters of the supercell of WSe$_2$ (8.7 Å) match well with the lattice parameters of EuS (8.44 Å).

The WSe$_2$ has a two dimensional hexagonal unit cell with the lattice parameter of 3.28 Å, while EuS has a cubic unit cell with the lattice parameter of 5.968 Å. In order to simulate their interface using DFT methods and take into account 2D periodicity of the system we constructed a supercell of WSe$_2$ on the slab of (111) surface of EuS. Considering the lattice parameters of (111) surface is 8.44 Å, i.e. incommensurate with the lattice parameter of WSe$_2$ of 3.28 Å. However, the supercell of $\vec{a'} = 3\vec{a} - \vec{b}$ ($\vec{a}, \vec{b}$ are primitive lattice vectors of WSe$_2$) as shown in Fig. 1b. The supercell contains 7 unit cells of WSe$_2$. Due to the different bond length of Se-Se in WSe$_2$ and in EuS there are possible choices of WSe$_2$ unit cell shift. We find that Eu sitting at the center of one of the sites threefold coordinated by Se is a stable one. Three other Eu at the surface are also coordinated by three Se atoms (but Eu position is shifted in the direction of one of the Se in the triangle). After relaxation these Eu Sites are located at the bridge positions between two Se atoms.

Because of the mismatch of the lattice parameters, we used WSe$_2$ lattice parameters in the calculations because WSe$_2$ is not expected to have a large strain after deposition. At the same time the main effects from the ferromagnetic substrate will be captured despite a slight bi-axial tensile strain.

The (111) surface is polar, which usually reconstructs with half of Eu sites (or S site in sulfur-terminated surface) missing. This would make pyramids of EuS with (001) facets.

The band structure of the supercell contains bands from both EuS and WSe$_2$. Majority of the bands are formed by the electrons of either EuS or WSe$_2$. There is a small overlap of orbitals for a few bands related to interface. The equilibrium Eu-Se distance at the interface is larger than 3.08 Å, i.e. relatively large for forming strong covalent bonds that could lead to band mixing. We plotted the bands of WSe$_2$ character to discuss the effect of the ferromagnetic interface on the optical transitions in this system.

The positions of top valence and bottom of the conduction bands at K and K' points of the Brillouin Zone are given in Table I. It clearly shows the splitting due to the exchange field produced by the EuS substrate.

Table I. Energy of the top valence and bottom conduction bands at K and K' for each spin.

| energies (eV) | Spin up | Spin down |
| --- | --- | --- |
| E(v, K) | 0.578 | 0.340 |
| E(c, K) | 1.684 | 1.602 |
| E(v, K') | 0.341 | 0.582 |
| E(c, K') | 1.579 | 1.698 |